\documentclass[aps,prb,showpacs,twocolumn]{revtex4}
\usepackage{graphicx}
\usepackage{dcolumn}
\usepackage{bm}
\usepackage[latin1]{inputenc}
\usepackage{amsmath}

\usepackage{colordvi}
\usepackage{color}
\usepackage{ulem}

\begin{document}

\title{Spin freezing by Anderson localization in one-dimensional semiconductors}

\author{C. Echeverr\'ia-Arrondo}
\affiliation{Department of Physical Chemistry, Universidad del Pa\'is Vasco
UPV/EHU, 48080 Bilbao, Spain} 
\author{E. Ya. Sherman}
\affiliation{Department of Physical Chemistry, Universidad del Pa\'is Vasco
UPV/EHU, 48080 Bilbao, Spain} 
\affiliation{IKERBASQUE Basque Foundation for Science, 48011 Bibao, Bizkaia, Spain} 

\date{\today}

\begin{abstract}
One-dimensional quantum wires are considered as prospective elements for spin transport
and manipulation in spintronics.  
We study spin dynamics in semiconductor GaAs-like nanowires with
disorder and spin-orbit interaction by using 
a rotation in the spin subspace gauging away the spin-orbit field. 
At a strong enough disorder spin density, 
after a relatively fast relaxation stage, reaches a plateau, 
which remains a constant for long time. This effect is a manifestation 
of the Anderson localization and depends in a universal
way on the disorder and the spin-orbit coupling strength. 
As a result, at a given disorder, semiconductor nanowires
can permit a long-term spin polarization tunable with the spin-orbit interactions.
\end{abstract}


\maketitle

\section{Introduction}

The main idea of spintronics - the design and 
application of devices controlling not only the charge
dynamics but also the electron spin evolution - can be useful 
for information storage, transfer, and manipulation 
technologies.\cite{spintronics1,spintronics2,MWWu} 
Possible realizations of spintronics
devices can be based on 
semiconductor nanowires\cite{nw1,Nadj,Pramanik,Kiselev,Quay,Bringer,Governale} for quasi-ballistic
electron transport, coherent transmission of information, and spin
control. These systems attract a great deal of attention due
to a clear interplay of transport and spin-orbit (SO) coupling
physics.\cite{Pershin,Holleitner,Sanchez,Lu,Romano,Japaridze}    

This control faces the problem of inevitable spin relaxation due to the coupling 
of electron spin to environment through SO coupling.  As a result, the factors
determining the spin relaxation rate become of crucial 
importance. Two limiting cases of spin 
relaxation are well understood. For the itinerant electrons spin relaxation in mainly
determined by the Dyakonov-Perel' mechanism, that is by random precession of electron
spin due to the random in time electron momentum.

A different approach should be applied for electrons localized in a regular 
external potential forming quantum dots promising for quantum information applications.\cite{LAWu} 
Here momentum is not a well-defined quantity, and the momentum-dependent splitting
required for the Dyakonov-Perel mechanism vanishes. As a result,  
spin relaxation through SO interaction requires phonon-induced coupling of different 
orbital states of the localized electron and nonzero external magnetic field.\cite{Stano}
In the absence of magnetic field and spin-orbit coupling, 
spin relaxation can occur due to the hyperfine coupling of electron
spin to spins of lattice nuclei.\cite{Merkulov} In both cases, the initial spin polarization 
goes asymptotically to zero. The characteristic 
timescale of spin relaxation of electrons localized in quantum dots is expected
to be several orders of magnitude longer than that of itinerant electrons.

\begin{figure}[tbp]
\includegraphics*[scale=.36]{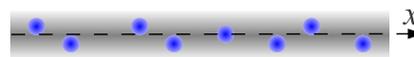}
\caption{(Color online) Semiconductor nanowire with random impurities shown as filled circles. 
Although we consider a one-dimensional electron motion, impurities can be randomly distributed over the
cross-section of the wire.}
\label{fig:fig1}
\end{figure}

While these two limits of free and strongly localized electrons 
are well understood, the interplay of disorder-induced 
localization and spin relaxation of itinerant electrons remains an open question
although some aspects of the problem have been addressed.\cite{Bruno,Shklovskii,Kaneko,Intronati}
The questions here are (i) how the localization forms the spin
relaxation, and (ii) whether the initially prepared spin density relaxes to zero.  
As a nontrivial example of this interplay we mention that weak localization of two-dimensional 
electrons leads to a long power-like rather than exponential spin relaxation.\cite{Lyubinskiy,Tokatly}
Here we analyze this problem for the one-dimensional system, providing, on one hand,
the basic example of localization physics in a random potential,\cite{Anderson,disorder}  
and, on the other hand, an example of a system, where spin-orbit coupling can be
gauged away by a SU(2) transformation. 

This paper is organized as follows. In Sec. II, we show how to treat spin relaxation in one-dimensional
systems with the gauge transformation and introduce the tight-binding Hamiltonian for the model.
The spin dynamics will be analyzed by a numerically exact calculation in Sec. III,
where we show that spin density does not relax to zero, in contrast to what expected. In addition,
in this Sec. III we study how asymptotic value of spin polarization depends on the disorder 
and spin-orbit coupling. Conclusions summarize the results in Sec. IV.   

\section{Model}

\subsection{Hamiltonian and gauge transformation}

The investigated structure is a quantum wire extended along the $x$ axis, as shown in Fig.~\ref
{fig:fig1}. 
The total Hamiltonian has the form
\begin{equation}
\hat{H}=\frac{\hbar^2}{2m}\left(k_{x}-A_{x}\right)^{2}+U(x)-\frac{m\alpha^{2}}{2\hbar^2},  \label{hso}
\end{equation}
where $A_{x}=-m\alpha\sigma_{y}/\hbar^{2}$ stands for the Rashba coupling \cite{Bychkov} 
with the strength $\alpha$, $\sigma_{y}$
is the Pauli matrix, $k_{x}$ is the electron wavevector, and $m$ is the 
effective mass. The Dresselhaus coupling \cite{dresselhaus1} is obtained with 
$A_{x}=-m\beta\sigma_{x}/\hbar^{2}$, where $\beta$ is the coupling constant. 
Without loss of generality, we concentrate here
on the Rashba coupling, which can be changed on demand by applying external electric 
field across the structure.\cite{Karimov03} 

The SO interaction can be removed from $\hat{H}$ 
in Eq.(\ref{hso}) through a gauge transformation \cite{LevitovRashba,gauge} 
with a SU(2) spin rotation: $\hat{S}=\exp\left(-ix\sigma_{y}/2\xi\right),$
where $\xi =\hbar^{2}/2m\alpha$ is the spin-precession length. 
After this transformation the system
Hamiltonian has the form:  $\hat{\widetilde{H}}={\hbar^2 k_{x}^{2}}/{2m}+U(x).$

Since for the Hamiltonian (\ref{hso}), $\sigma_{y}$ 
is the integral of motion, the spin density component along the $y$-axis is time independent. 
A nontrivial dynamics of the transformed spin occurs for the $\gamma=(x,z)$ spin components
$\langle\widetilde{s}_{\gamma}\left( x,t\right)\rangle$ and can be expressed in terms
of the spin diffusion 
\begin{equation}
\langle\widetilde{s}_{\gamma }\left( x,t\right)\rangle =
\int{D}^{\gamma \beta }(x-x^{\prime },t)
\langle\widetilde{s}_{\beta }\left( x^{\prime },0\right)\rangle dx^{\prime },
\label{S:evolution}
\end{equation}
where ${D}^{\gamma\beta}(x,t)$ is the exact disorder-dependent one-dimensional spin
diffusion Green's function. In a nonmagnetic system without SO coupling  
${D}^{\gamma \beta }(x,t)=\delta _{\gamma \beta }{D}(x,t)$ is diagonal in the spin subspace.
As a result of the gauge transformation, the uniform density dynamics 
is determined by only the Fourier component\cite{Tokatly}  
\begin{equation}
{D}(q,t)=\int_{-\infty}^{\infty} dxe^{-iqx}{D}(x,t)  \label{D:q}
\end{equation}
with $q=1/2\xi$ and Eq.~(\ref{S:evolution}) simplifies for the physical measurable 
spins as $\langle{s}_{\gamma}\left(t\right)\rangle
=\langle{s}_{\gamma}\left(0\right)\rangle{D}(1/2\xi,t)$. 
Here we will use a similar,
however, somewhat different approach based on numerically exact analysis of the direct
time evolution of the initial spin-polarized states. 
It will be shown that the resulting spin dynamics 
has unexpected features, including a long-time plateau in the spin polarization. 

\begin{figure}[tbp]
\includegraphics[scale=0.5]{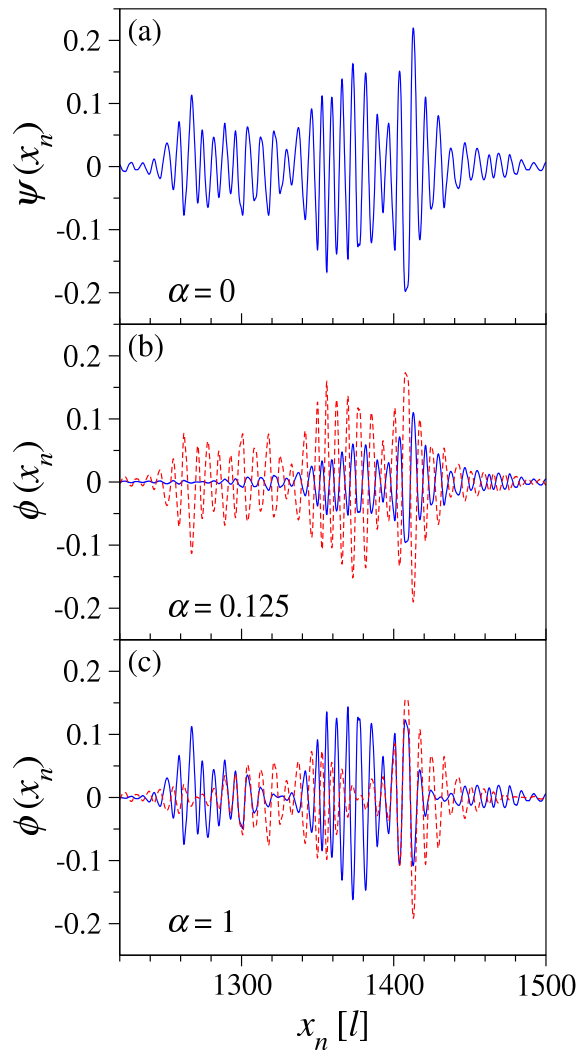}
\caption{(Color online) Site dependent components of $\overline\phi_{N/4}(x_n)$
for a qualitative description of entanglement induced by the gauge
transformation for (a) $\alpha=0$, (b) $\alpha=0.125\times10^{-6}$ meVcm, and (c)
$\alpha=10^{-6}$ meVcm ($U_{0}=55$ meV).  The solid and dashed lines represent $\left|1\right>$ and $\left|-1\right>$ 
components, respectively.}
\label{fig:fig2}
\end{figure}

The eigenfunctions of $\hat{\widetilde{H}}$ can 
be taken in the form $\overline{\psi}(x)={\psi}(x)\left|1\right\rangle$
and $\overline{\psi}(x)={\psi}(x)\left|-1\right\rangle$, where 
$\left|\pm1\right\rangle$ are the eigenstates of $\sigma_z$ with the corresponding eigenvalues. 
The eigenstates of $\hat{H}$, ${\overline{\phi}}(x)$ can be obtained by spin
rotation of the $\overline{\psi}(x)\left|\sigma\right>$ states. For example,
with spin-up initial state $\overline{\psi}(x)\left|1\right>$ one obtains: 
\begin{equation}
\overline{\phi}(x)=\psi(x)
\left[
\cos\left(\frac{x}{2\xi}\right)\left|1\right\rangle
+\sin\left(\frac{x}{2\xi}\right)\left|-1\right\rangle
\right].  \label{ent}
\end{equation}
The spin dynamics and spin relaxation in the system, as
it will be shown below, is solely due to the entanglement of spin and coordinate
in Eq.(\ref{ent}). 

\subsection{Tight-binding model and disorder}

We perform numerical analysis using the tight-binding
model, employing the approach similar to Refs.[\onlinecite{Japaridze,Hankiewicz}]. 
The one-dimensional electron gas is sampled with $N=2^{13}$ (8192) 
grid points $x_{n}=nl$, where $1\le n\le N$ and $l$ is the effective 
lattice constant with periodic boundary conditions.\cite{parpack}
The effective hopping matrix element between two nearest neighbors
is chosen as $t=50$~meV, and the kinetic energy is $E(k_{x})=2t(1-\cos(k_{x}l))$. 
As a result, the eigenenergies span the range of $[0,200]$~meVs.
The distance between two neighbor grid points becomes $l=\hbar/\sqrt{2mt}=3.37$ nm to satisfy
the electron effective mass $m=0.067~m_{0}$ in GaAs semiconductor
with $m_{0}$ being the free electron mass. 

The random potential $U_{n}=U(x_{n})$ uniformly spans the range $[-U_{0}/2,U_{0}/2]$ with the white noise
correlator $\left\langle U(x_{n_{1}})U(x_{n_{2}})\right\rangle =\langle
U^{2}\rangle \delta_{n_{1},n_{2}}$, where $\langle U^{2}\rangle=U_{0}^{2}/12$. 
The effects of disorder can be approximately characterized through the energy-dependent 
momentum relaxation time $\tau_{E}$, which we define as $\hbar/\tau_{E} =\langle U^{2}\rangle l\nu_{E} $, where
$\nu_{E} =\sqrt{m}/\pi\hbar\sqrt{2E}$ is the density of states per spin component. 
The resulting mean free path $\ell_{E}=v_{E}\tau_E$, where $v_{E}=\sqrt{2E/m}$ is the 
electron speed and the corresponding diffusion coefficient $D_E=v_{E}^{2}\tau_{E}$. 

In this representation the eigenstates of $\hat{\widetilde{H}}$ and $\hat{H}$  form basis sets, 
$\{\overline{\psi}_{i}\}$ and $\{\overline{\phi}_{i}\}$ respectively, where $1\le i\le 2N$. For the
same $i$, these two sets are related by the local spin rotation $\hat{S}.$ We assume that
$\overline{\psi}_{i}={\psi}_{i}(x_{n})\left|1\right>$ for $1\le i\le N$ and 
$\overline{\psi}_{i}={\psi}_{i-N}(x_{n})\left|-1\right>$ for $N<i\le 2N.$ 

\section{Spin dynamics}

We study dynamics of initial $\overline{\psi}_{i}$ states with $1\le i\le N$, 
corresponding to the evolution upon instant switching of the SO coupling. 
The time dependence can be expressed with the spectral decomposition as:
\begin{equation}
\overline{\psi}_{j}^{\mathrm{so}}(t)=
\sum_{i=1,2N}a_{ij}\overline{\phi}_{i}e^{-it\varepsilon_{i}/\hbar},
\end{equation}
where $a_{ij}=\langle \overline{\phi}_{i}|\overline{\psi}_{j}\rangle$, and $\varepsilon_i$ 
are the corresponding eigenenergies. The spin component expectation value
$\left\langle \sigma _{z}(t)\right\rangle_{j} =
\left\langle\overline{\psi}_{j}^{\mathrm{so}}(t)\right|\sigma_{z}\left|\overline{\psi}_{j}^{\mathrm{so}}(t)\right\rangle$ 
is determined by the spectrum and eigenstates of the system. 

\begin{figure}[tbp]
\includegraphics[scale=0.45]{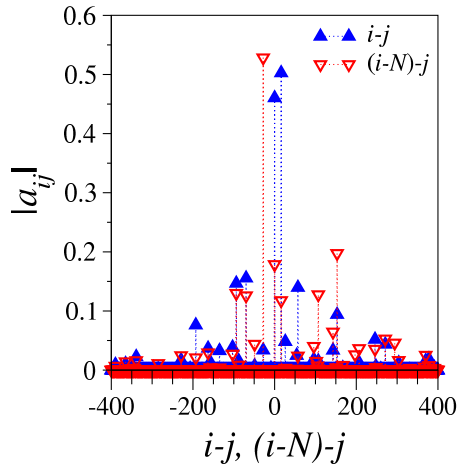}
\caption{(Color online) Absolute values of $a_{ij}$ around the initial spin-up state $\overline\psi_{N/4}$; 
here $\alpha=10^{-6} \mbox{ meV}\mbox{cm}$ (strong SO coupling) 
and $U_0$=55~meV (strong disorder).}
\label{fig:fig3}
\end{figure}

In order to give an idea of the entanglement induced by 
SO coupling, we present in Fig.~\ref{fig:fig2} the evolution 
of $\overline{\phi}_{N/4}(x_{n})$ state with the increase in the
spin-orbit coupling. At $\alpha=0$, we obtain a product state
$\overline{\phi}_{N/4}(x_{n})=\psi_{N/4}(x_{n})\left|1\right>$, and 
with the increase in $\alpha$ entangled states are formed. 
The overlap of $\overline{\phi}_{i}(x_{n})$ and $\overline{\psi}_{j}(x_{n})$ 
eigenstates is characterized by two sets of matrix 
elements $a_{ij}$; for example, for $1\le j\le N:$
\begin{eqnarray} 
&&a_{ij}=\sum_{n}\cos\left(\frac{x_{n}}{2\xi}\right)\psi_{i}(x_{n})\psi_{j}(x_{n}),\quad 1\le i\le N\\
&&a_{ij}=\sum_{n}\sin\left(\frac{x_{n}}{2\xi}\right)\psi_{i}(x_{n})\psi_{j}(x_{n}),\quad N<i\le 2N.\nonumber
\label{aij-previous}
\end{eqnarray}
The behavior of $a_{ij}$ presented Fig.~\ref{fig:fig3} demonstrates that for given $j$
it has nonnegligible values only in a certain, rather narrow, range of $i$. 

\begin{figure}[tbp]
\includegraphics*[scale=0.45]{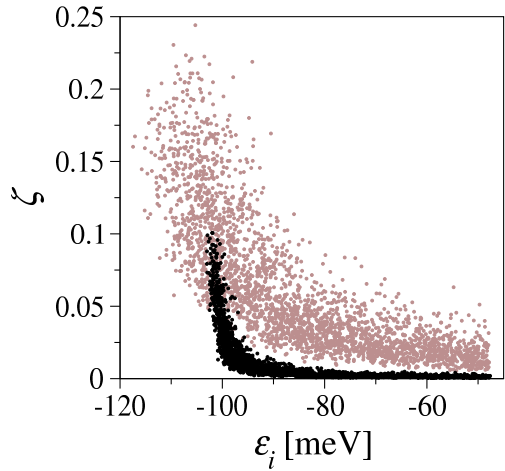}
\caption{(Color online) Inverse participation ratio $\zeta$ for the low
part of the energy spectrum; gray (red) circles denote
strong disorder ($U_0$=55~meV) and black circles denote weak disorder ($U_0$=15~meV).
Since even for $U_0$=55~meV we obtain $\zeta\ll 1$, the localized states are distributed
over many lattice sites, confirming applicability of the tight-binding Hamiltonian for the localization
problem.}
\label{fig:fig4}
\end{figure}

\begin{figure}[tbp]
\includegraphics[scale=0.4]{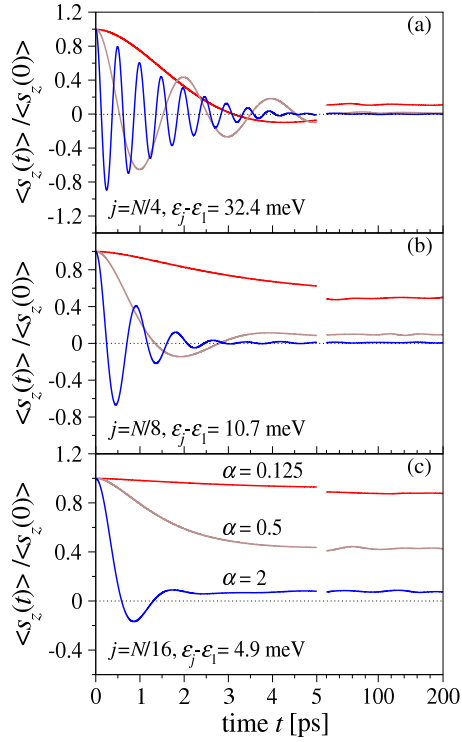}
\caption{(Color online) Time-dependent polarization in the weak-disorder regime 
($U_0$=15~meV). The initial bins are centered at the states (a) $N/4$ (bin width 6.8 meV),
(b) $N/8$ (bin width 3.7 meV), and (c) and $N/16$ (bin width 2.1 meV) with energies decreasing in
the same order. The curves for SO couplings $0.125\times 10^{-6}\mbox{ meV}\mbox{cm}$, 
$0.5\times 10^{-6} \mbox{ meV}\mbox{cm}$, and $2\times 10^{-6}\mbox{ meV}\mbox{cm}$ are
drawn with circles, triangles, and squares, respectively. Note that after the relaxation stage the spin density
remains a finite constant.}
\label{fig:fig5}
\end{figure}

\begin{figure}[tbp]
\includegraphics[scale=0.4]{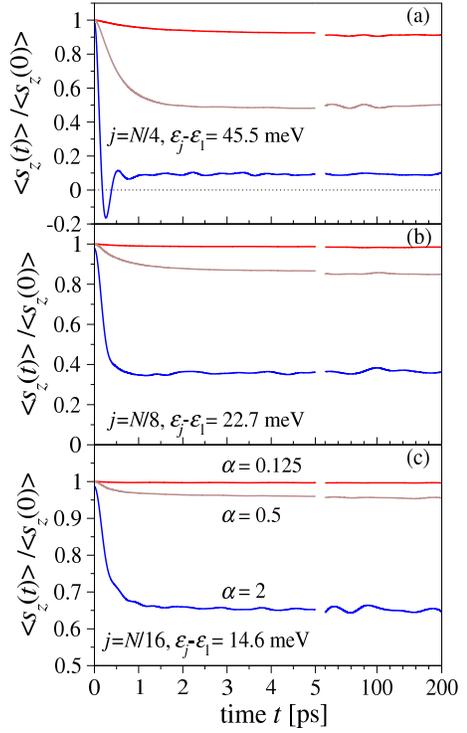}
\caption{(Color online) Time-dependent spin polarization in the strong-disorder regime
($U_0$=55~meV) with the same notations as in Fig.~\ref{fig:fig5}. (a) $N/4$ (bin width 7.2 meV), 
(b) $N/8$ (bin width 4.5 meV), and (c) $N/16$  (bin width 3.7 meV). Note that
for $\alpha=0.125\times 10^{-6} \mbox{ meV}\mbox{cm}$ the spin is almost constant in
time, thus suitable for spin-based operations.}
\label{fig:fig6}
\end{figure}

To illustrate the role of the random potential, 
we consider as examples weak ($U_{0}=15$ meV, $U_{0}\ll t$) and 
strong ($U_{0}=55$ meV, $U_{0}>t$) disorder. 
For free electrons in state $j=N/4$ and $E=31$ meV, 
the resulting $\hbar/\tau_{E}$ is about 0.1 and 1~meV, respectively. 
For a free electron with the energy $E\approx 20$ meV the velocity 
$v_{E}\approx 3.5\times10^{7}$ cm/s, the mean free path $\ell_{E}\sim 2.5\times10^{-5}$ cm 
($\hbar/\tau_{E}=1$ meV), and the corresponding diffusion coefficient $D_E\sim10^{3}$ cm$^{2}$/s. 
These parameters provide an effective integral characteristic of the disorder 
and correspond to realistic parameters of the wires, which, however, can strongly vary 
from sample to sample and from experiment to experiment.

The effect of localization by disorder is seen in the inverse participation ratio \cite{IPR} (IPR) 
$\zeta_{i}=\sum_{n}\left|\psi_{i}^{4}(x_{n})\right|$. The IPR 
calculated for the low-energy spectrum is presented in Fig.~\ref{fig:fig4}. 
As expected, the degree of localization increases with $U_{0}$ 
and this effect is more pronounced for the electrons with lowest energies.
In contrast to the results of Ref.[\onlinecite{Intronati}], the IPR in this system
does not depend on the SO coupling. 
We now study the effects of disorder and spin-orbit coupling on the average spin dynamics of a bin of 256 
initial spin-up states and 8 realizations of the 
random potential. The statistical error of this approach is, 
therefore $1/\sqrt{2048}$=2.2\%, making the results statistically
representative. 

We take three example bins with three different
degrees of localization. The bins are centered around the spin-up states $%
\overline\psi_{N/4}$, $\overline\psi_{N/8}$, and $\overline\psi_{N/16}$, whose IPR
values increase in the same order (energies decreasing, see Fig.~\ref
{fig:fig4}). The calculated bin- and potential realization-averaged spin dynamics is
shown in Figs.~\ref{fig:fig5} and \ref{fig:fig6}, revealing
strong influence of the disorder-induced spatial localization of states. Physically, 
collisions of electrons with impurities force electron spin to frequently reverse
the precession direction. In the classical picture,
this leads to a long Dyakonov-Perel'  spin relaxation. 
If the quantum effects of localization are important, the 
resulting effect is the ``freezing'' of the
electronic spin. As one can see in Figs.~\ref{fig:fig5} and \ref{fig:fig6}, the electron
spin density relaxes for $\simeq5$~ps and then remains constant in time for
infinitely long (beyond 0.2~ns in our computation). As expected, the spin polarization 
plateau is higher (i) for localized states and (ii) for weak SO interaction. Almost 
time-independent spin states are achieved e.g., 
at $U_0=55$~meV and $\alpha=0.125\times 10^{-6} \mbox{ meV}\mbox{cm}$.

\begin{figure}[t]
\includegraphics*[scale=0.36]{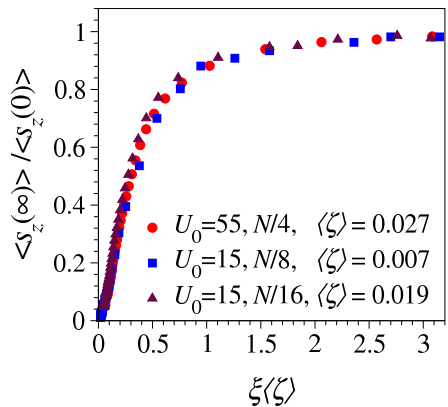}
\caption{(Color online) Long-term relative polarization as a function of 
$\xi\langle\zeta\rangle$ for three different degrees of localization. Parameter $\xi$ 
is modified by changing the coupling constant $\alpha$.}
\label{fig:fig7}
\end{figure}

To gain insight into the problem, we study the dependence of asymptotic spin density on SO coupling  and the 
localization of electrons in more detail. The 
long-term densities are plotted in Fig.~\ref{fig:fig7} against parameter $\xi \langle
\zeta \rangle $. This parameter combines the two factors determining the spin
dynamics, SO coupling  and spatial localization, where $\langle \zeta \rangle $ is
averaged over 256 bin states and 8 realization of the random potential. The given values
follow a universal dependence indicating  a unique trend for long-term
spins against SO coupling  and localization through disorder. This trend 
corresponds to a fast increase in the asymptotic steady polarization for $\xi \langle
\zeta \rangle <1$ and a smooth increase and saturation for $\xi \langle \zeta \rangle >1$.
These results can be understood as follows. To show an efficient
spin dynamics, the electron should move the distance of the order of $\pi\xi$.
Therefore, the spatial spread of the corresponding states should be larger than 
$\pi\xi$. With a stronger localization, the spread and the overlap decrease
leading to the universal behavior shown
in Fig.\ref{fig:fig7}. Qualitatively, in the  ``clean''  $\xi\langle\zeta\rangle\ll 1$ regime the spin
relaxation has the Dyakonov-Perel' mechanism either purely 
exponential for $\Omega_{E} \tau_{E}\ll1$ or a combination
of oscillations and exponential decay if $\Omega_{E}\tau_{E}\ge1$, where the spin precession
rate $\Omega_{E} =2\alpha\sqrt{2m{E}}/\hbar$ corresponds to the electron momentum at given
energy $E$.

\section{Conclusion}

To summarize, localization effects of disorder and SO coupling in semiconductor nanowires
determine the dynamics of electronic spins. Our tight-binding model
calculations show that a prepared spin density relaxes until reaching a
plateau, directly related to the disorder and strength of SO interaction. 
In contrast to the expected decay to zero, a long-time constant polarization 
plateau survives to infinite time.
The asymptotic spin density has a universal dependence on the product of the inverse participation
ratio and the spin precession length.   
In the absence of magnetic field, the hyperfine coupling to the spins of nuclei 
will lead to spin relaxation on timescales at least two orders of magnitude
longer than the timescale of the plateau formation of the order of 10 ps.\cite{Merkulov}  
As the experiments on spin transport did not reveal electron-electron interaction effects,\cite{Quay} here we have neglected them. Furthermore, whether there exists 
a range of parameters where the Coulomb forces can be strong enough to modify our results for localized states, remains to be investigated.  

An immediate consequence of this result is the ability, by choosing the desired Rashba
SO parameter for a given wire, to produce and destroy steady spin states,
which are of interest for spin-based operations. These results
suggest that semiconductor nanowires can be used for coherent transmission  
and storage of information, manipulated by spatially and temporally 
modulated spin-orbit coupling. 

\section{Acknowledgments}

We thank G. Japaridze, J. Siewert, and L.A. Wu  for helpful
discussions. This work was supported by the 
MCI of Spain grant FIS2009-12773-C02-01 and "Grupos Consolidados UPV/EHU 
del Gobierno Vasco" grant IT-472-10.


\end{document}